\documentclass[11pt]{article}
\usepackage{hyperref}
\pdfoutput=1 
\usepackage{titling}
\begin{document}
\title{Bouncing, Helical and Buckling Instabilities \\ During Droplet Collision:\\ Newtonian and Non-Newtonian Liquids}
\author{Xiaodong Chen and Vigor Yang \\
\\\vspace{6pt} School of Aerospace
Engineering, \\ Georgia Institute of Technology, Atlanta, GA 30332, USA}
\maketitle
While collision of low viscous Newtonian liquid droplets has been extensively investigated both experimentally and numerically for decades, limited is known about viscous Newtonian and non-Newtonian droplet collision dynamics.  In the present work related to this fluid dynamics video, high-fidelity numerical simulations were performed to study the situation associated with a viscous Newtonian liquid and a shear-thinning non-Newtonian liquid. The viscosity of the Newtonian fluid is set to be equal to the maximum equivalent viscosity of the non-Newtonian liquid. The formulation is based on a complete set of conservation equations for both the liquid and the surrounding gas phases.  An improved volume-of-fluid (VOF) method, combined with an innovative topology-oriented adaptive mesh refinement (TOAMR) technique was developed and implemented to track the interfacial dynamics.  The complex evolution of the droplet surface over a broad range of length scales was treated accurately and efficiently. Especially, the thin gas film between two approaching droplets and subsequent breakup of liquid thread were well-resolved. Regime diagrams were developed and compared for the two types of liquids. Fundamental mechanisms and key parameters that dictate droplet behaviors were identified.

In this video, Ray-tracing data visualization technique was used to obtain realistic and detailed flow motions during droplet collision. The differences of collision outcome between Newtonian and non-Newtonian were compared. Various types of droplet collision were presented, including bouncing, coalescence, and stretching separation. Because of the reducing of equivalent viscosity caused by shear stress, the gas film between shear-thinning droplet is thinner than Newtonian liquid. Since thinner gas film promotes coalescence, shear thinning liquid has smaller area of bouncing regime in the diagram of Weber number and impact parameter. During the ligament/thread breakup process of stretching separation, two kinds of instabilities are identified, helical and buckling instabilities. Helical instability is analogous to a viscous rotating liquid jet, while the buckling instability is analogous to electrically charged liquid jets of polymer solutions.
\end{document}